\pdfoutput=1
\documentclass[fp,twocolumn]{jpsj3}
\usepackage{txfonts}
\usepackage{bm}
\usepackage{color}

\newcommand{\SO}{S\!O}

\title{Possibility of a Topological Phase Transition in Two-dimensional $RP^3$ Model}

\author{Tsuyoshi Okubo$^{1}$\thanks{t-okubo@phys.s.u-tokyo.ac.jp} and Naoki Kawashima$^{2,3}$}
\inst{$^1$ Institute for Physics of Intelligence, University of Tokyo, Tokyo 133-0033, Japan\\
$^2$ Institute for Solid State Physics, University of Tokyo, Kashiwa, Chiba 277-8581, Japan\\
$^3$ Trans-scale Quantum Science Institute,
University of Tokyo, Tokyo 113-0033, Japan} 

\date{\today}

\abst{We study by large-scale Monte Carlo simulation the $RP^3$ model, which can be regarded as an effective low-energy model of a triangular lattice Heisenberg antiferromagnet. $Z_2$ vortices appear as elementary excitations in the triangular lattice Heisenberg antiferromagnet. Such $Z_2$ vortices are ubiquitous in other frustrated Heisenberg spin systems that have noncollinear long-range orders. In this study, we investigate a possible topological phase transition driven by the binding--unbinding of $Z_2$ vortices. By extracting important degrees of freedom, we map a frustrated spin system to an effective $RP^3$ model. From large-scale Monte Carlo simulation, we obtain an order parameter and a correlation length of up to $L=16384$. Concerning the existence of a $Z_2$-vortex transition, by extrapolating the order parameter to the thermodynamics limit assuming the $Z_2$-vortex transition, we obtain a finite transition temperature as $T_v/\tilde{J} \simeq 0.25$. Our estimate of the correlation length at $T_v$ is much larger than $L=16384$, which is beyond the previous estimate obtained with the triangular lattice Heisenberg model.
}
\begin{document}

\maketitle
\section{Introduction}\label{sec:introduction}
Frustrated interactions in spin systems induce various properties in systems \cite{Diep2004,LacroixMM2011,Balents2010}. For example, in several frustrated systems, classical ground states are macroscopically degenerated; in such cases, the systems do not exhibit a magnetic long-range order even at zero temperature. When the quantum fluctuation is introduced to such systems, typically in the case of $S=1/2$ quantum spins, their ground states could be quantum spin liquid, as we see in Kagom\'{e} lattice or pyrochlore lattice Heisenberg antiferromagnets \cite{LacroixMM2011,Balents2010, LiaoXCLXHNX2017,GingrasMC2014}.

\begin{figure}[th] 
 \begin{center} 
  \includegraphics[width=0.9\linewidth]{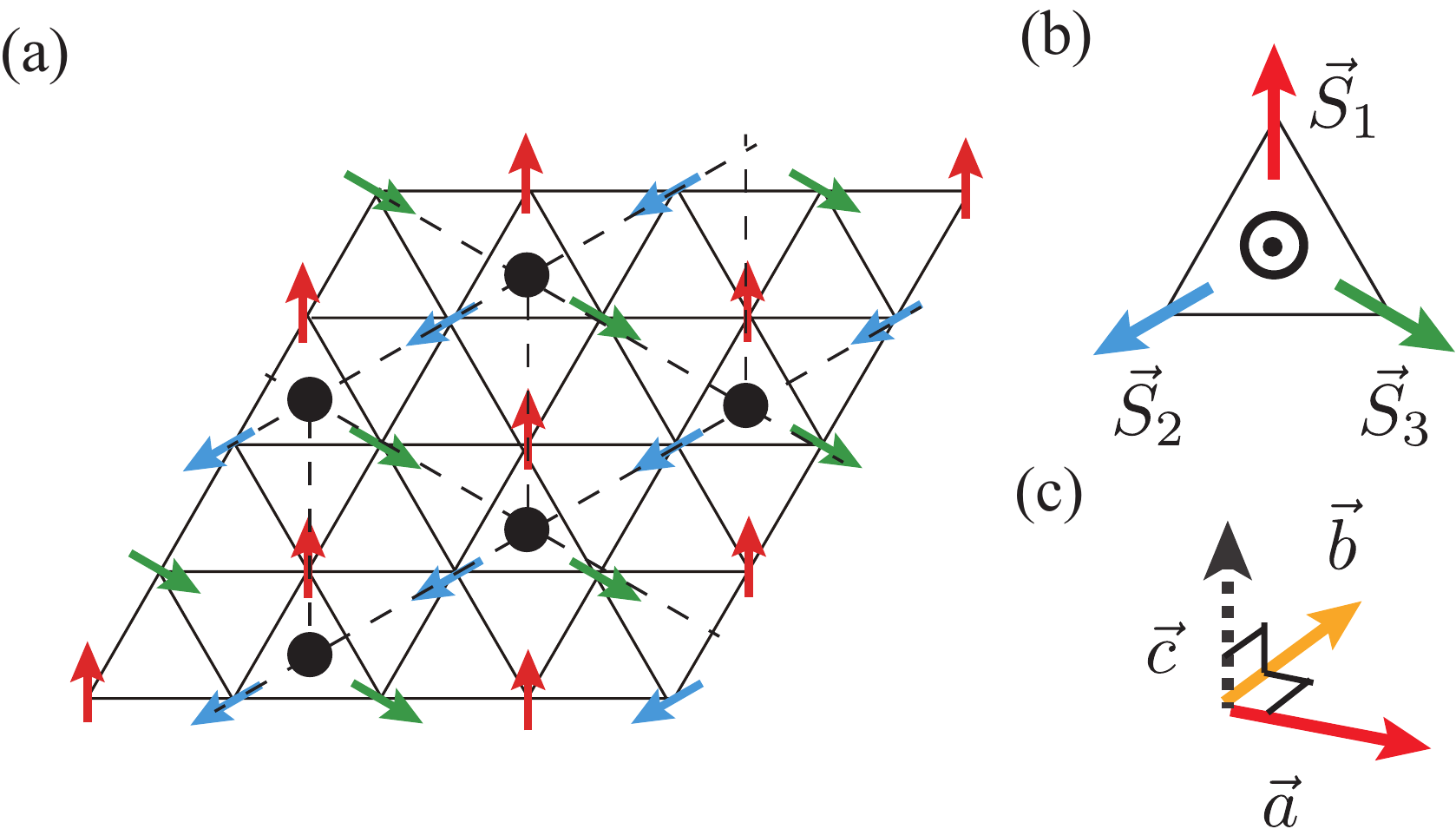}
 \end{center} 
 \caption{(Color online) (a) Schematic view of 120 degree structure on the triangular lattice. Black circles and dotted lines indicate the coarse-grained triangular lattice formed by upward triangles. (b) Local 120 degree structure together with the vector chirality $\kappa \propto \vec{S}_1 \times S_2$ perpendicular to the plane. (c) Local orthonormal  vectors $\vec{a}, \vec{b}$, and $\vec{c}$ defined from a local 120 degree structure.} \label{fig:120_structure}
\end{figure}

Another important aspect of the frustration involves noncollinear and noncoplanar spin correlations. Owing to the competition among interactions, even if a system exhibits a magnetic long-ranger order at low temperatures, spins tend to cant each other. Such canted spin textures induce a new degree of freedom in addition to the original spins. A well-known example is the triangular lattice $XY$ model; its ground state is the 120 degree structure as shown in Fig.~\ref{fig:120_structure}(a). In this case, the ground state manifold is characterized by $O(2)$, and then, in addition to standard the $U(1)$ spin rotation, a new $Z_2$ degree of freedom characterized by the vector chirality $\vec{\kappa} = \vec{S}_i \times \vec{S}_j$ appears. It was argued that there exist two finite temperature phase transitions corresponding to $Z_2$ and $U(1)$ symmetry breaking. Here, the phase transition related to $U(1)$ spin rotation is a type of Berezinskii--Kosterlitz--Thouless (BKT) transition explained by the binding--unbinding of vortices characterized by winding numbers indexed by integers. Such vortices are sometimes called $Z$ vortices \cite{Berezinsky1970,KosterlitzT1973}.

When we look at the Heisenberg model on the triangular lattice, we observe even more interesting properties. Although the ground state of the Heisenberg model has again the 120 degree structure, its order parameter space is characterized by $\SO(3)$ instead of $O(2)$ of the XY model. Similarly to the XY model, a topological excitation plays an important role in disturbing the magnetic long-range order, while its topological nature is different from the $Z$ vortex. The vortex excitation in the Heisenberg model is characterized by $Z_2$ indices instead of $Z$. A possible topological phase transition driven by $Z_2$ vortices was proposed by Kawamura and Miyashita nearly four decades ago \cite{KawamuraM1984}. One of the important predictions for the $Z_2$-vortex transition is that the spin correlation length remains finite at the transition temperature $T_v$. At $T_v$, the vortex correlation length characterizing a typical distance between free vortices diverges, while the spin correlation length is finite even below $T_v$ \cite{KawamuraYO2010, OkuboK2010}. This is a crucial difference from the well-known BKT transition, in which both the spin correlation and vortex correlation lengths diverge at the transition temperature \cite{Berezinsky1970,KosterlitzT1973}. Note that the topological nature depends only on the symmetry of the order parameter space. Thus, we expect the $Z_2$-vortex transition in various frustrated Heisenberg models on two-dimensional lattices.

Similar topological phase transitions have also been discussed in liquid crystal systems \cite{KunzZ1989,KunzZ1992,LebwohlL1972,Tomita2014,LathaS2018}. A recent Monte Carlo (MC) simulation showed that two topological phase transitions with nondivergent correlation lengths occur in a two-dimensional nematic three-vector model \cite{LathaS2018}. One of these transitions might be related to the $Z_2$-vortex transition discussed in the frustrated Heisenberg models.

The existence of the $Z_2$-vortex transition in the triangular lattice Heisenberg antiferromagnet has been investigated mainly using MC simulations \cite{KawamuraM1984, KawamuraYO2010, SouthernX1995,WintelEA1995}. Previous MC simulations for $L \times L$ triangular lattices up to $L=1536$ showed that the $Z_2$-vortex transition occurs at $T_v/J = 0.285$ \cite{KawamuraYO2010}. However, the nature of the $Z_2$ vortex transition has not been fully understood because the estimated spin correlation length at $T_v$ is $\xi \simeq 2000$, which is comparable to the largest system size. Under such a situation, one might consider that the observed $T_v$ is just a crossover due to the finite size effect. Here, we address this concern by extending the system size range on the order of magnitude. As we discuss below, the best estimate of the transition temperature (assuming it is finite) is updated to a lower value, and the new estimate of the correlation length at the criticality is about one order of magnitude larger than the previous estimate. Therefore, in this paper, as well, we do not observe the case where the correlation length fits well within the system size. This may sound that we are suggesting the absence of the finite-temperature phase transition. However, our fitting assuming the finite-temperature phase transition is better than that assuming the zero-temperature transition. Although we will not draw a clear yes/no answer to the central question, we will set a new and more stringent condition for a finite-temperature transition. 

Larger-scale simulations of frustrated spin models are usually difficult. Since cluster update techniques, such as Swendsen--Wang and Wolff algorithms, do not work efficiently, MC simulations suffer from a critical slowing down; the typical relaxation time increases as a function of system size as
\begin{equation}
 \tau \propto L^z,
\end{equation}
with $z > 2$. This means that to double the system size, we must pay for the time cost of MC simulations that is at least 16 times larger for the same accuracy. Such a huge computation cost has limited the typical simulation size to $L \lesssim 5000$ and clarifying the existence of the $Z_2$-vortex transition has been difficult so far.

In this study, to overcome such difficulties, we investigate an effective model for the $Z_2$ vortex transition that represents interactions among local $\SO(3)$ degrees of freedom \cite{CaffarelADM2001,KawamuraK1993}. When the $Z_2$ vortex transition exists in the thermodynamic limit, it must be independent of details of systems, and it is expected to exist in any model that keeps the same symmetry as the original triangular lattice Heisenberg model. As we see later, by considering the effective model, we can apparently eliminate the frustrated interactions, and this allows the Swendsen--Wang cluster algorithm to work more efficiently than in the case of the original triangular lattice model.

The rest of the paper is organized as follows. In Sec.~\ref{sec:model}, we introduce an effective model for the $Z_2$-vortex transition and describe a numerical method to investigate the model. In Sec.~\ref{sec:results}, we show the results obtained by large-scale MC simulations for the effective model, and we discuss the nature of the $Z_2$-vortex transition on the basis of results. Section \ref{sec:conclusion} is devoted to the conclusions.

\section{$RP^3$ Model and Its Relationship with the Frustrated Heisenberg Antiferromagnet}
\label{sec:model} 
In this paper, we exclusively study the $RP^3$ model. Below, we show how this model is related to previously studied models in the context of the $Z_2$-vortex transition.

To investigate the possibility of the $Z_2$-vortex transition, we consider an effective model for the low-temperature properties of frustrated Heisenberg models, {\it e.g.}, the triangular lattice antiferromagnetic Heisenberg model \cite{CaffarelADM2001,KawamuraK1993}. The Hamiltonian of the triangular lattice antiferromagnetic Heisenberg model is given by
\begin{equation}
 \mathcal{H} = J \sum_{\langle i,j\rangle} \vec{S}_i\cdot \vec{S}_j,
\end{equation}
where $\vec{S} = (S_x, S_y, S_z)$ is a three-component unit vector, $J > 0$, and $\sum_{\langle i,j\rangle}$ means the sum over the nearest-neighbor pairs on the triangular lattice. 

The ground state of the model is the $120$ degree structure [Fig. \ref{fig:120_structure}(a)]. To construct the effective model, here we focus on local $120$ degree structures defined on (upward) triangles. Suppose that all local $120$ degree structures are rigid. In this situation, the remaining degrees of freedom are their \textit{directions}. Because the direction of a local rigid $120$ degree structure can be characterized by an $\SO(3)$ matrix, we can define the direction by using three orthonormal vectors. We can construct the orthonormal vectors $\vec{a}$, $\vec{b}$, and $\vec{c}$ from three spins on a triangle:
\begin{align}
 \vec{a} &= \vec{S}_1 \notag \\
 \vec{c} &\propto \vec{S}_1 \times \vec{S}_2 \notag \\
 \vec{b} &= \vec{c} \times \vec{a},
\end{align}
 [see Figs.~\ref{fig:120_structure}(b) and \ref{fig:120_structure}(c)]. By using these vectors, we obtain the low-temperature effective model as
\begin{equation}
 \mathcal{H}_{\mathrm{eff}} = -\sum_{\langle i,j\rangle_{t}}\left(p_1 \vec{a}_i\cdot  \vec{a}_j + p_1 \vec{b}_i\cdot  \vec{b}_j + p_3 \vec{c}_i\cdot  \vec{c}_j\right),
 \label{eq:H_eff}
\end{equation}
where a set of three vectors $(\vec{a}_i,\vec{b}_i,\vec{c}_i)$ are located at an upward triangle of the original triangular lattice and $\sum_{\langle i,j\rangle_{t} }$ is the sum over the nearest neighbors on the coarse-grained triangular lattice formed by upper triangles [see Fig.~\ref{fig:120_structure}(a)]. Note that the interaction coefficients $p_1$ and $p_3$ are non-negative (ferromagnetic) such that the ground state corresponds to the original $120$ degree structure. The coefficient $p_3$ is generally different from $p_1$ because $\vec{c}_i$ is perpendicular to local $120$ degree structures, whereas $\vec{a}_i$ and $\vec{b}_i$ are on the plane where the original spins form a local $120$-degree structure. The study conducted by Kawamura and Kikuchi corresponds to the model with $p_3=0$ \cite{KawamuraK1993}. Note that the effective model contains only ferromagnetic interactions, and there is no frustration. Therefore, the underlying lattice structure is irrelevant to the nature of ordering. For simplicity, hereafter, we investigate the model on the square lattice, instead of the original triangular lattice. 

By introducing a $3\times 3$ matrix $R_i$ as $R_i = (\vec{a}_i,\vec{b}_i, \vec{c}_i)$, we can transform the Hamiltonian \eqref{eq:H_eff} into a simpler form as
\begin{equation}
 \mathcal{H}_{\mathrm{eff}} = -\sum_{\langle i,j\rangle} \mathrm{Tr}~R_i^t P R_j,
\end{equation}
where $P$ is a diagonal matrix $P=\mathrm{diag}~(p_1,p_1,p_3)$. From this representation, we can see that the Hamiltonian is unchanged under the transformation
\begin{equation}
 R_i' = U R_i V,
\label{eq:trans}
\end{equation}
where $U$ is an $\SO(3)$ rotational matrix and $V$ is an $O(2)$ rotational matrix that mixes $\vec{a}$ and $\vec{b}$. Because of these symmetries, the present effective model is often called the $\SO(3) \times O(2)$ model. When we set $p_3=p_1$, the model becomes $\SO(3) \times \SO(3)$ symmetric, {\it i.e.}, the Hamiltonian is unchanged under the transformation \eqref{eq:trans} with both $U$ and $V$ being $\SO(3)$ matrices. 

Because the low-temperature states of the effective model have topological defects, $Z_2$ vortices, independent of $p_3$, here, we concentrate on the case of $p_3=p_1$, {\it i.e.}, the  $\SO(3) \times \SO(3)$ model. The $\SO(3) \times \SO(3)$ model can be mapped onto the so-called $RP^3$ model \cite{CaffarelADM2001}. In the $RP^3$ model, the {\it four-component} unit vectors $\vec{S}_i=(S_{i,0},S_{i,1},S_{i,2},S_{i,3})$, interact through the biquadratic form as
\begin{equation}
 \mathcal{H}_{RP^3} = -\tilde{J} \sum_{\langle i,j\rangle} \left(\vec{S}_i\cdot\vec{S}_j\right)^2.
\label{eq:H_RP3}
\end{equation}
An effective spin $\vec{S}_i$ in the $RP^3$ model is related to the matrix $R_i$ through the relation
\begin{multline}
 R_{i,kl} = 2\left(S_{i,k}S_{i,l} - \frac{1}{4}\delta_{k,l}\right) + 2\sum_{m=1}^3 \epsilon_{klm} S_{i,0}S_{i,m}\\ + 2\left(S_{i,0}^2 -\frac{1}{4}\right)\delta_{kl}.
\end{multline}

One can also consider the $RP^{n-1}$ model, which is a generalization of the $RP^3$ model. In the $RP^{n-1}$ model, $n$-component spins interact with each other through the biquadratic interactions. Such an $RP^{n-1}$ model has been investigated, {\it e.g.}, as a model for liquid crystals \cite{KunzZ1989,KunzZ1992,LebwohlL1972,Tomita2014,LathaS2018}. In the case of $n=2$, the model is equivalent to the XY model and exhibits the BKT transition on two-dimensional lattices. For $n\ge 3$, two-dimensional $RP^{n-1}$ models have $Z_2$ vortices and could exhibit the $Z_2$-vortex transition as we expect for the $RP^3$ model. In the limit of $n \to \infty$, the model becomes soluble: the model exhibits a finite-temperature first-order phase transition \cite{KunzZ1992}.

Note that by changing the model from the triangular lattice Heisenberg model to the $RP^3$ model, the effective system size is increased because $\SO(3)$ matrix $R_i$ represents a local 120 degree structure consisting of three spins. Thus, we expect the $RP^3$ model on the $L \times L$ lattice to correspond to the triangular lattice Heisenberg model larger than $L \times L$. This is an advantage of investigating the effective model instead of the original triangular lattice Heisenberg model. 

To investigate the $RP^3$ model on the $L\times L$ square lattice numerically, we conducted MC simulations. Because the model contains only ferromagnetic interactions and there is no frustration, one might expect that standard cluster algorithms work efficiently. Kunz and Zumbach have proposed a Wolff--Swendsen--Wang-type cluster algorithm for general $RP^{n-1}$ models \cite{KunzZ1989} and it has been extended to the $\SO(3)\times O(2)$ model by Caffarel {\it el al.} \cite{CaffarelADM2001}. In the algorithm, we first generate a random $n$-dimensional unit vector $\vec{e}$ and then calculate the projection of $\vec{S}_i$ onto the $\vec{e}$ as $\sigma_i = \vec{S}_i\cdot \vec{e}$. Finally, we perform the Swendsen--Wang cluster algorithm \cite{SwendsenW1987} by constructing clusters defined using effective Ising variables $\sigma_i$.

For small system sizes, the Wolff--Swendsen--Wang algorithm for the $RP^{n-1}$ and $\SO(3) \times O(2)$ models has worked efficiently \cite{KunzZ1989,CaffarelADM2001}. However, for systems of $L=O(1000)$, we found that the relaxation of $Z_2$ vortices suffers from a kind of critical slowing down, and then the total relaxation time increases as $\tau \propto L^z$ with $z\simeq 2$. Thus, for the present model, the cluster algorithm works not surprisingly well, although it can largely reduce the relaxation time from that of local updates.

To investigate much larger system sizes than those in the previous simulations, we also employed the MPI parallelization of the Wolff--Swendsen--Wang algorithm. We split the two-dimensional square lattice into $M \times M$ cells, and an MPI process is assigned to each cell. By using such a massively parallelized Wolff-Swendsen-Wang algorithm, we performed MC simulations of the square lattice $RP^3$ model up to $L = 16384$. In the vicinity of the expected $Z_2$ vortex transition, the relaxation time becomes longer. We observed the slowest relaxation at around $T/\tilde{J} = 0.28$, and at that temperature, we performed MC simulations with $4.8 \times 10^7$ MC steps for $L=16384$ and $9.6 \times 10^7$ MC steps for $L=8192$; for other temperatures and smaller $L$s, we carried out shorter MC simulations. The statistical error of physical quantities was estimated from 3--4 independent runs. 

\section{Results}
\label{sec:results} 
\begin{figure}[th] 
 \begin{center} 
  \includegraphics[width=0.9\linewidth]{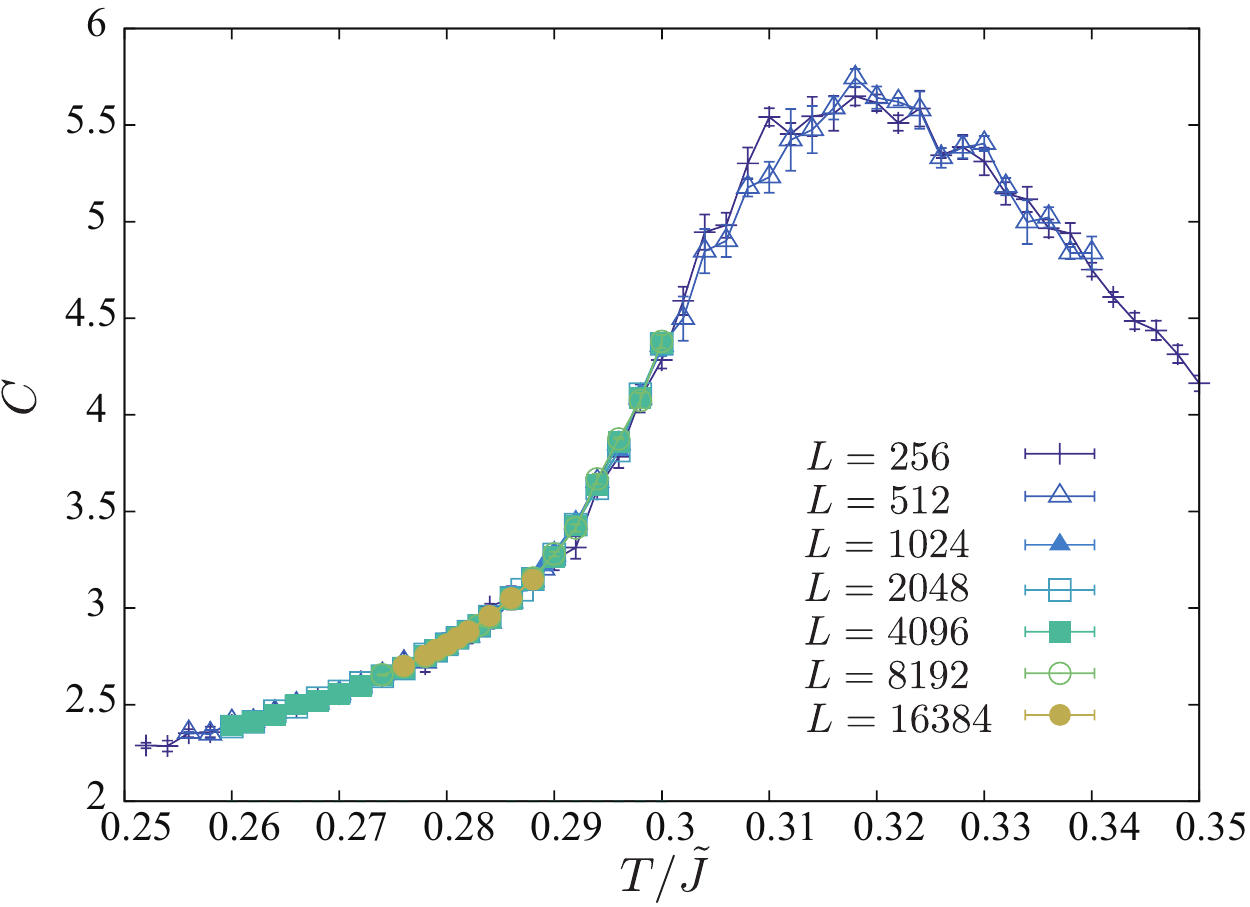}
 \end{center} 
 \caption{(Color online) Temperature dependence of specific heat for various system sizes. } \label{fig:specific_heat}
\end{figure}

\begin{figure}[th] 
 \begin{center} 
  \includegraphics[width=0.9\linewidth]{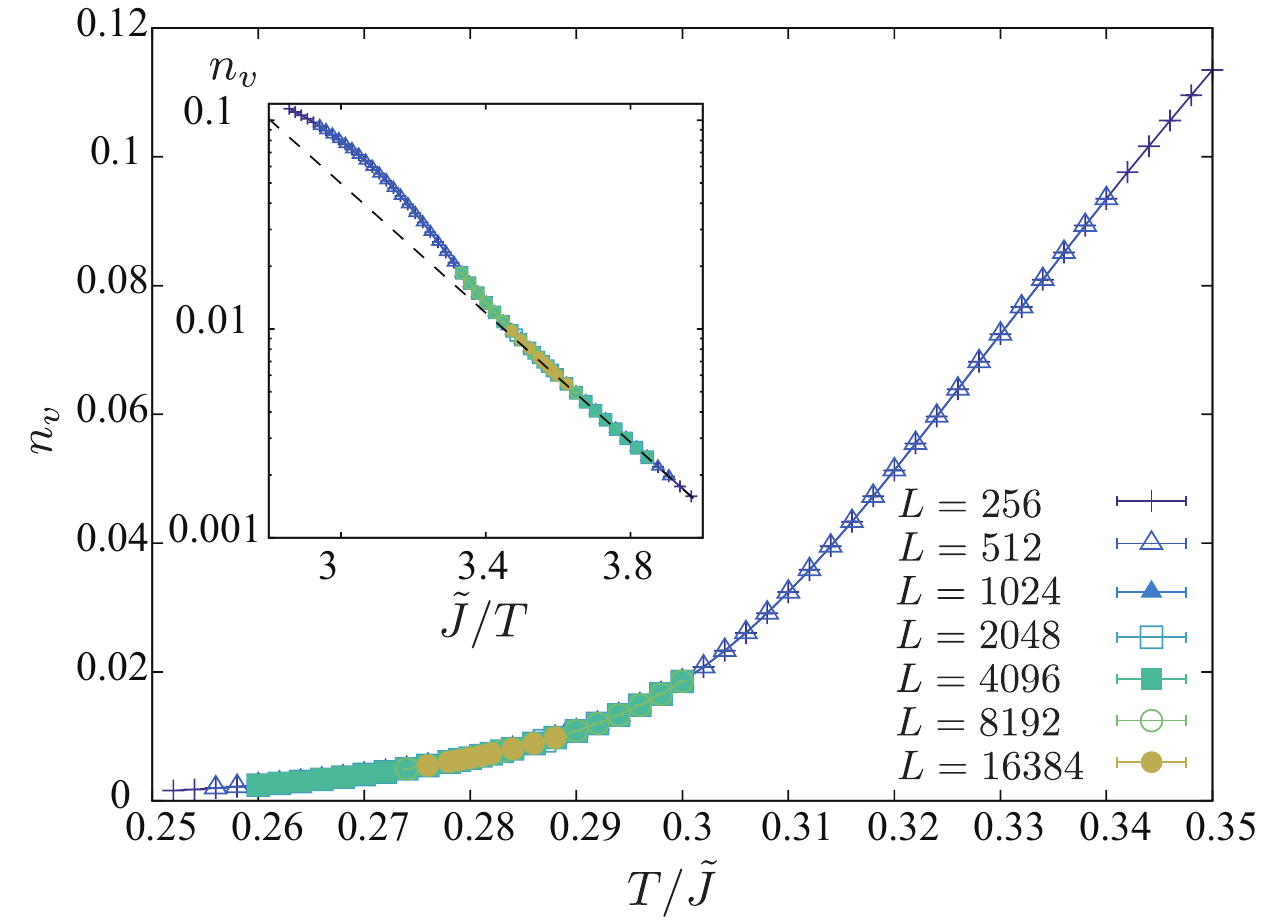}
 \end{center} 
 \caption{(Color online) Temperature dependence of vortex density. The inset shows the vortex density in the semi-log plot as a function of $1/T$. The dashed curve indicates the fitting with $n_v \propto e^{-\mu_c/T}$. } \label{fig:vortex_density}
\end{figure}

Figure~\ref{fig:specific_heat} shows the specific heat for various $L$s. As we expect for the phase transition with the essential singularity \cite{KawamuraYO2010}, there is a broad peak corresponding to the development of the short-range magnetic (nematic) order, {\it i.e.,} we do not observe a clear singularity in the specific heat. To see the behaviors in the vortex sector, we show the temperature dependence of $Z_2$-vortex density in Fig.~\ref{fig:vortex_density}. We define $Z_2$ vortices on every elemental plaquette of the square lattice as defined in Ref.~\citen{KunzZ1992}: where we assign $\pm 1$ to each bond that connects sites $i$ and $j$ by the $\mathrm{sign}~\vec{S}_i\cdot \vec{S}_j$ and put a vortex if the product of the signs on four edges surrounding the plaquette is $-1$. The vortex density $n_v$ is defined as the ratio of the number of vortices to the volume $L^2$. As we see in Fig.~\ref{fig:vortex_density}, $n_v$ does not show a strong size dependence. Because $n_v$ contains both free vortices and vortex pairs, it is difficult to observe a singular behavior related to the possible $Z_2$-vortex transition associated with the binding--unbinding of the free vortices \cite{KawamuraM1984,KawamuraYO2010}. Nevertheless, we can also realize that $n_v$ rapidly varies at around $T/\tilde{J}\simeq 0.28$. We might understand it through the thermal activation of vortex pairs below the $Z_2$-vortex transition. In the low-temperature phase, the vortices appear only as pairs and the vortex density is expected to be described by the thermal activation of such pairs, $n_v\propto e^{-\mu_c/T}$, where $\mu_c$ is twice the chemical potential of a vortex. The inset in Fig.~\ref{fig:vortex_density} shows $n_v$ as a function of $\tilde{J}/T$ in the semi-log plot. For $T/\tilde{J} \gtrsim 0.28 $ ($\tilde{J}/T \lesssim 3.6$), we see a clear deviation from the behavior $n_v\propto e^{-\mu_c/T}$ with $\mu_c \simeq 3.57$, indicating that at least something happens in the vortex degrees at around $T/{\tilde{J}} \simeq 0.28$. 

Note that $T/{\tilde{J}} \simeq 0.28$ is close to the estimated $Z_2$-vortex transition temperature $T_{v} \simeq 0.285$ in the triangular lattice Heisenberg model \cite{KawamuraYO2010}. This is probably related to the observation that $\mu_c \simeq 3.57$ is close to that of the triangular lattice Heisenberg model, $\mu_c \simeq 3.3$ \cite{KawamuraM1984} , indicating that the RP3 model has a similar energy scale to the triangular lattice Heisenberg model.

To clarify the existence of the $Z_2$-vortex transition, here we consider the vorticity modulus as an order parameter of the topological phase transition \cite{KawamuraK1993,SouthernX1995}. The vorticity modulus is defined by the free energy difference between the systems with and without a free vortex\cite{KawamuraK1993}. Owing to the spin modulation caused by a vortex, the free energy difference $\Delta F$ is expected to have logarithmic size dependence for a sufficiently large $L$ as
\begin{equation}
 \Delta F = C + v \log L,
\label{eq:V}
\end{equation}
and the coefficient $v$ in front of  $\log L$ is the vorticity modulus. In the thermodynamic limit, the vorticity modulus is zero ($v=0$) for $T > T_v$ and is finite ($v > 0$) for $T < T_v$. Thus, we can determine the transition temperature $T_v$ from the size and temperature dependences of the vorticity modulus.

In our MC simulations, we calculate the vorticity modulus through the standard equilibrium samplings as in Refs.~\citen{KawamuraYO2010}, \citen{SouthernX1995}, and \citen{KawamuraY2007}. By assuming a simple vortex structure, we can calculate the free energy difference at a finite size $L$, $\Delta F(L)$, using the spin fluctuation in the equilibrium configurations under the periodic boundary condition \cite{SouthernX1995};it is similar to the helicity modulus, which is often used in the investigatation of the BKT transition \cite{SouthernY1993}. Using the language of the $\SO(3)$ model, we consider a single vortex structure formed by $\vec{a}_i$, $\vec{b}_i$ and $\vec{c}_i$ rotating around a certain axis in a three-dimensional spin space. Then, we can extract the vorticity modulus $v$ from $\Delta F(L)$s of two different system sizes $L_1$ and $L_2$ as
\begin{equation}
 v(L_1,L_2) \equiv \frac{\Delta F(L_1) - \Delta F(L_2)}{\log L_1/L_2}.
\label{eq:vorticity}
\end{equation}
By using the thus-defined vorticity modulus up to $L=1536$, Kawamura \textit{et al.} have estimated the $Z_2$-vortex transition temperature as $T_v \simeq 0.285\pm 0.005$ in the case of the triangular lattice Heisenberg model\cite{KawamuraYO2010}. Note that it does not largely deviate from $T_v \simeq 0.31\pm 0.01$ estimated from the Wilson loop type order parameter, which directly captures the binding--unbinding nature of the $Z_2$ vortices, calculated up to $L=36$\cite{KawamuraM1984}.

\begin{figure*}[th] 
 \begin{center} 
  \includegraphics[width=0.9\linewidth]{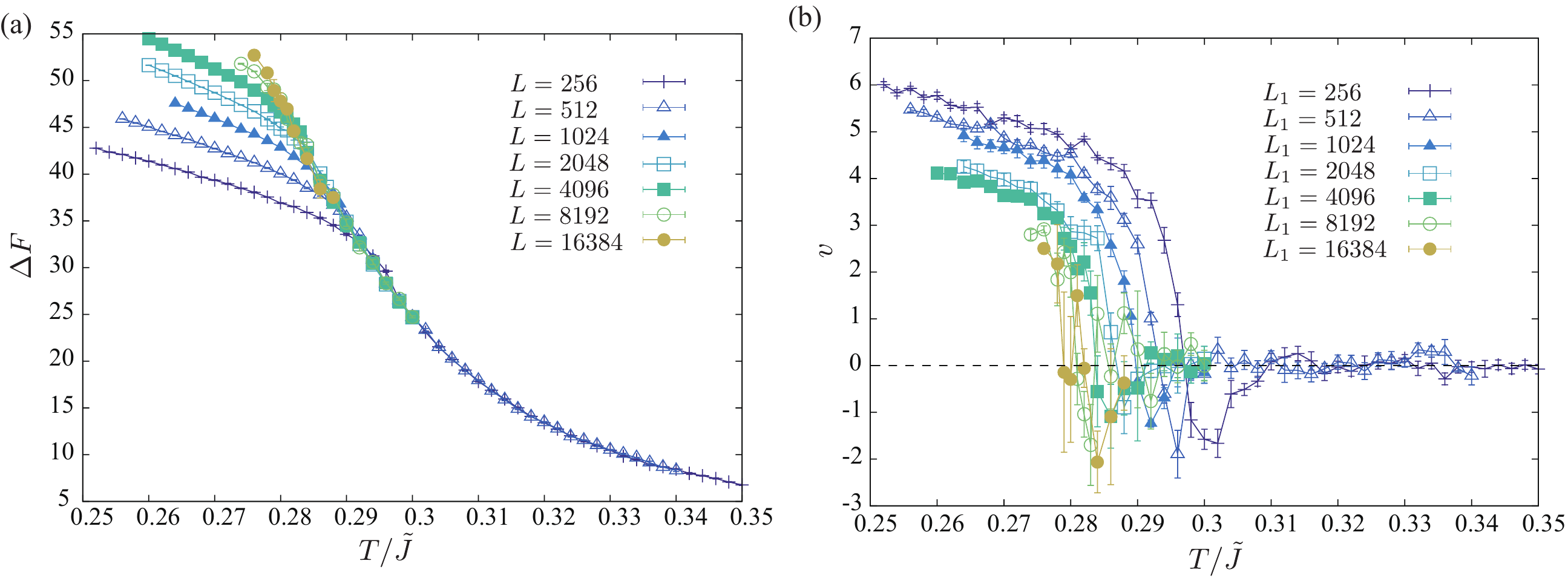}
 \end{center} 
 \caption{(a) (Color online) Free energy difference obtained by MC simulations for $128 \le L \le 16384$. (b) The vorticity modulus was calculated using Eq.~\eqref{eq:vorticity} by setting $L_2 = L_1/2$. The horizontal dashed line indicates $v=0$. }
 \label{fig:vorticity}
\end{figure*}

In Fig.~\ref{fig:vorticity}(a), we show the free energy difference $\Delta F(L)$ for various system sizes. As we decrease the temperature, $\Delta F$ increases: it corresponds to the increase in the free energy cost for generating a vortex. For $T \gtrsim 0.29 \tilde{J}$, $\Delta F(L)$ with $L \ge 256$ does not show a significant system size dependence, and then the vorticity modulus must be zero. In contrast, for lower temperatures, we can see a larger $L$ dependence of $\Delta F(L)$, indicating a finite vorticity modulus. From $\Delta F(L)$s of two system sizes $L_1$ and $L_2 = L_1/2$, we calculate the vorticity modulus using Eq.~\eqref{eq:vorticity}. Figure \ref{fig:vorticity}(b) shows the temperature dependence of the vorticity modulus for $256 \le L_1 \le 16384$. $v \simeq 0$ at sufficiently high temperatures indicates the proliferation of the free $Z_2$ vortices. When we decrease the temperature, \textit{e.g.,} $T \lesssim 0.3 \tilde{J}$, $v$ becomes finite, although we observe a strong system size dependence of $v$. Thus, to confirm the existence of the topological phase transition, we need to extrapolate $v$ to the thermodynamic limit. Note that $v$ becomes slightly negative upon cooling before it turns positive. It might be understood as a finite size effect from, {\it e.g.}, the $L$ dependence of the $C$ term in Eq.~\eqref{eq:V}. From this negative region, one can easily define the characteristic temperature $T_v(L)$ as the temperature where $v$ changes its sign. If the $Z_2$-vortex transition exists, $T_v(L)$ is expected to converge to a finite value in the thermodynamic limit, and the converged value must be identical to the transition temperature $T_v$. Note that $T_v(L)$ corresponds to the crossing point of the $\Delta F$ for $L_1$ and $L_2$. Thus, it might also be reasonable to define $T_v(L)$ as the characteristic temperature for $\Delta F$.

\begin{figure}[th] 
 \begin{center} 
  \includegraphics[width=0.9\linewidth]{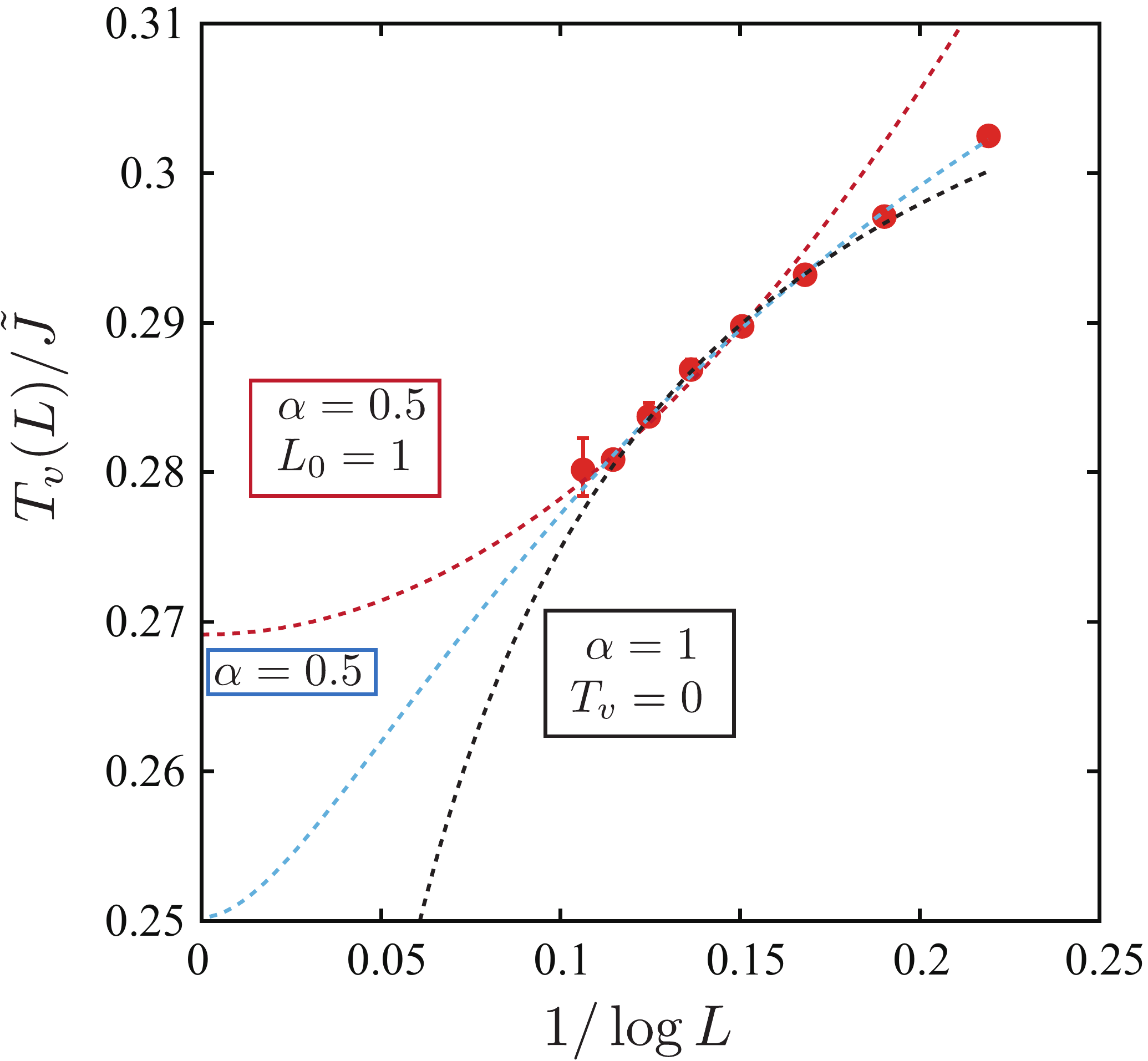}
 \end{center} 
 \caption{(Color online) Extrapolation of characteristic temperature $T_v(L)$ to the
 thermodynamics limit. Using Eq.~\eqref{eq:Tv_extra}, we extrapolated
 the data by assuming three situations: (red) $(\alpha=0.5,
 L_0=1)$, (blue) $(\alpha=0.5)$, and (black) $(\alpha=1, T_v=0)$.}
 \label{fig:tv_extra}
\end{figure}

In Fig.~\ref{fig:tv_extra}, we show $T_{v}(L=(L_1+L_2)/2)$ as a function of system size. To extrapolate the finite $T_{v}(L)$ to the thermodynamic limit, here we assume that the locations of the crossing points follow the finite-size scaling, which yields  
\begin{equation}
 T_v(L) = T_v + a \left[\log(L/L_0)\right]^{-\frac{1}{\alpha}},
\label{eq:Tv_extra}
\end{equation}
as proposed in Ref.~\citen{KawamuraYO2010}. This logarithmic function is natural because the correlation length of vortices, $\xi_v$, which is defined as a characteristic separation length of the free vortices, is expected to diverge exponentially toward $T_v$ as
\begin{equation}
 \xi_v \propto \exp\left[ \frac{A}{(T-T_v)^\alpha}\right],
  \label{eq:correlation}
\end{equation}
for $T > T_v$ \cite{KawamuraYO2010}.

In the case of the mean-field approximation, the exponent $\alpha$ is equal to $1$; on the other hande, by taking into account the fluctuations, $\alpha < 1$ is expected in the true $Z_2$-vortex transition \cite{KawamuraYO2010}. In the previous MC simulations of the triangular lattice antiferromagnetic Heisenberg model, $\alpha$ was estimated as $0.42 \pm 0.15$ \cite{KawamuraYO2010}, and it is not far from the value of the BKT transition, $\alpha = 0.5$. Note that even if the $Z_2$-vortex transition does not occur at a finite temperature, Eqs.~\eqref{eq:correlation} and \eqref{eq:Tv_extra} are applicable by setting $T_v=0$ and $\alpha=1$ because a renormalization group calculation of the corresponding nonlinear sigma model predicts the exponential growth of the spin correlation length $\xi$ toward $T=0$ \cite{AzariaDM1992}:
\begin{equation}
 \xi \propto T^{1/2}\exp\left[ \frac{A}{T}\right].
 \label{eq:no_Z2}
\end{equation}

Using Eq.~\eqref{eq:Tv_extra}, we try to extrapolate $T_v(L)$ to the thermodynamic limit by assuming $\alpha = 0.5$. Note that the exponent $\alpha = 0.5$ is different from that of a scenario without the $Z_2$-vortex transition ($\alpha = 1$). Thus, the fitting with $\alpha = 0.5$ somehow corresponds to assuming the existence of a finite-temperature $Z_2$-vortex transition. We also consider two situations for the parameter $L_0$: in one case, $L_0=1$ is assumed, and in the other case, we change $L_0$ freely. Note that in the previous analysis of the triangular lattice antiferromagnetic Heisenberg model, $L_0=1$ was used \cite{KawamuraYO2010} without explicitly mentioning the assumption. Although the fitting with $L_0=1$ seemed to work well in the previous analysis, as we see later, the choice of $L_0$ can largely change the extrapolation in the present $RP^3$ model.

In Fig.~\ref{fig:tv_extra}, we show the thus-obtained fitting curves. Both of the two curves with $\alpha=0.5$ give us the finite $T_v$s: They are $T_v/\tilde{J} \simeq 0.269$ and $T_v/\tilde{J} \simeq 0.250$ for $L_0=1$ and freely changed $L_0$, respectively. Therefore, when we assume the existence of the $Z_2$-vortex transition, we indeed obtain finite transition temperatures. Because there is no reason to choose $L_0=1$, $T_v/\tilde{J} \simeq 0.250$ is likely to be more reliable. 
In addition to these fitting curves, we also plot a fitting curve with assuming $\alpha=1$ and $T_v=0$ in Eq.~\eqref{eq:Tv_extra}: this corresponds to assuming without $Z_2$-vortex transition at a finite temperature. Although the fitting is not poor except for the largest size $L=16384$, we see that the fitting curve assuming $\alpha = 0.5$ fits over a wider range.
This finding could support the occurrence of the $Z_2$-vortex transition at a finite temperature.

To further clarify the possibility of the $Z_2$-vortex transition, here we investigate the correlation length with respect to the ``spin'' degrees of freedom. Note that in the present model, the original spins are coarse-grained as $\vec{a}$ and $\vec{b}$ in the $\SO(3) \times \SO(3)$ model and the original antiferromagnetic spin correlation corresponds to the ferromagnetic correlation among $\vec{a}$s and $\vec{b}$s. In the language of the $RP^3$ model, it is the ferronematic correlation of $\vec{S}_i$ of Eq.~\eqref{eq:H_RP3}. In the scenario of the finite-temperature $Z_2$ vortex transition, the correlation length does not diverge at the transition point and remains finite in the thermodynamic limit. As we discussed in Sec.~\ref{sec:introduction}, to obtain conclusive evidence of the $Z_2$-vortex transition numerically, the correlation length at the estimated transition temperature should be smaller than the maximum system size. Otherwise, it will be difficult to rule out the possibility that we are looking at merely a transient behavior and the extrapolation is not reliable owing to some finite-size effect not being taken into account.

\begin{figure}[th] 
 \begin{center} 
  \includegraphics[width=0.9\linewidth]{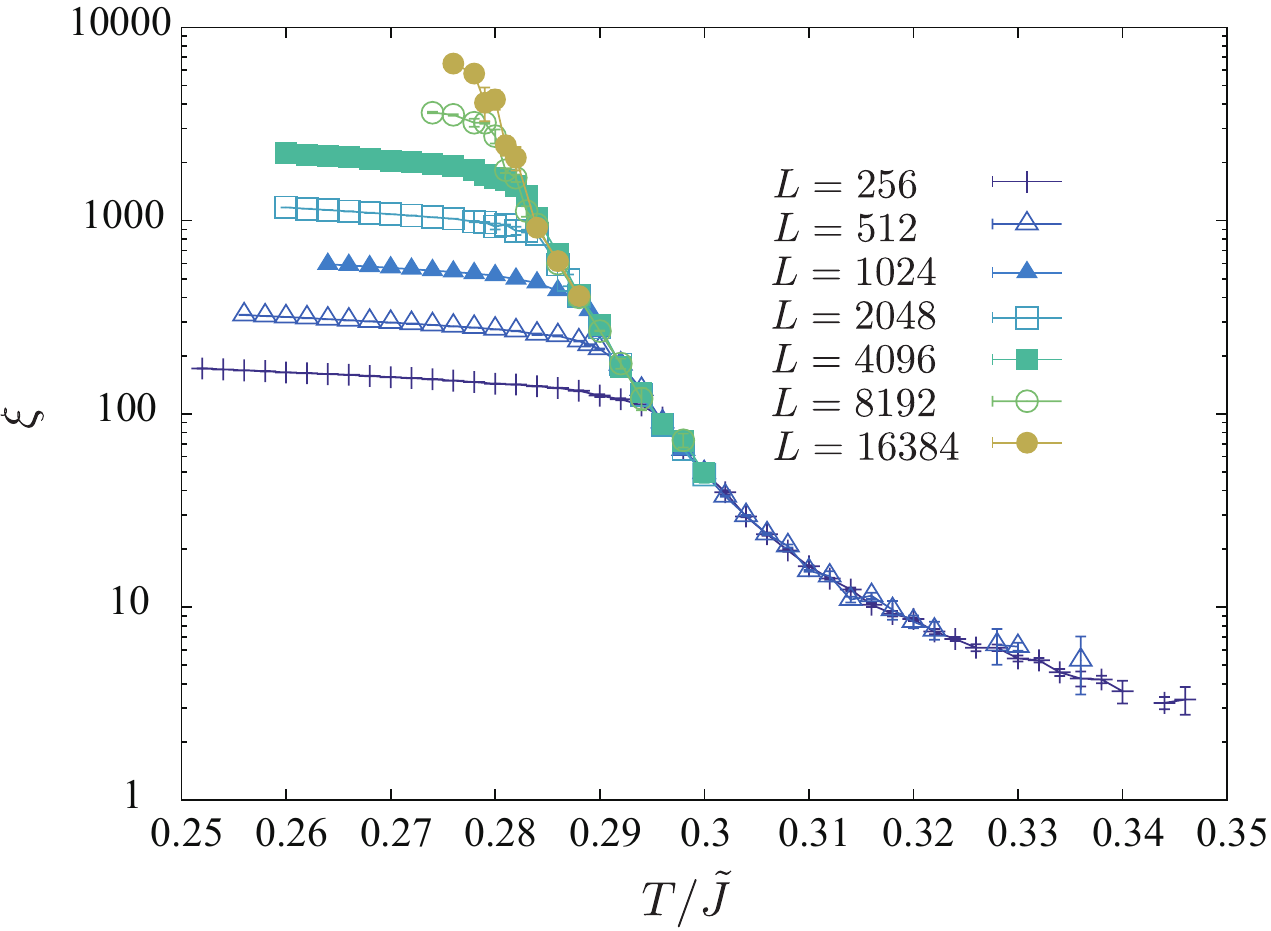}
 \end{center} 
 \caption{(Color online) The ``spin'' correlation length of $RP^3$ model for various system
  sizes.}  \label{fig:correlation}
\end{figure}

In Fig.~\ref{fig:correlation}, we show the correlation length of the model for various system sizes. We see that even at a higher estimated transition temperature, $T_v/\tilde{J}=0.269$, the correlation length seems to be larger than the largest system size of the present simulation, $L=16384$. Thus, we conclude that when the $Z_2$-vortex transition occurs, the correlation length at the transition temperature is much larger than the previous estimate, $\xi \simeq 1700$ \cite{KawamuraYO2010}. This also indicates that we need to investigate much larger systems to observe the separation of the two-length scales characterizing the $Z_2$-vortex transition.

\begin{figure}[th] 
 \begin{center} 
  \includegraphics[width=0.9\linewidth]{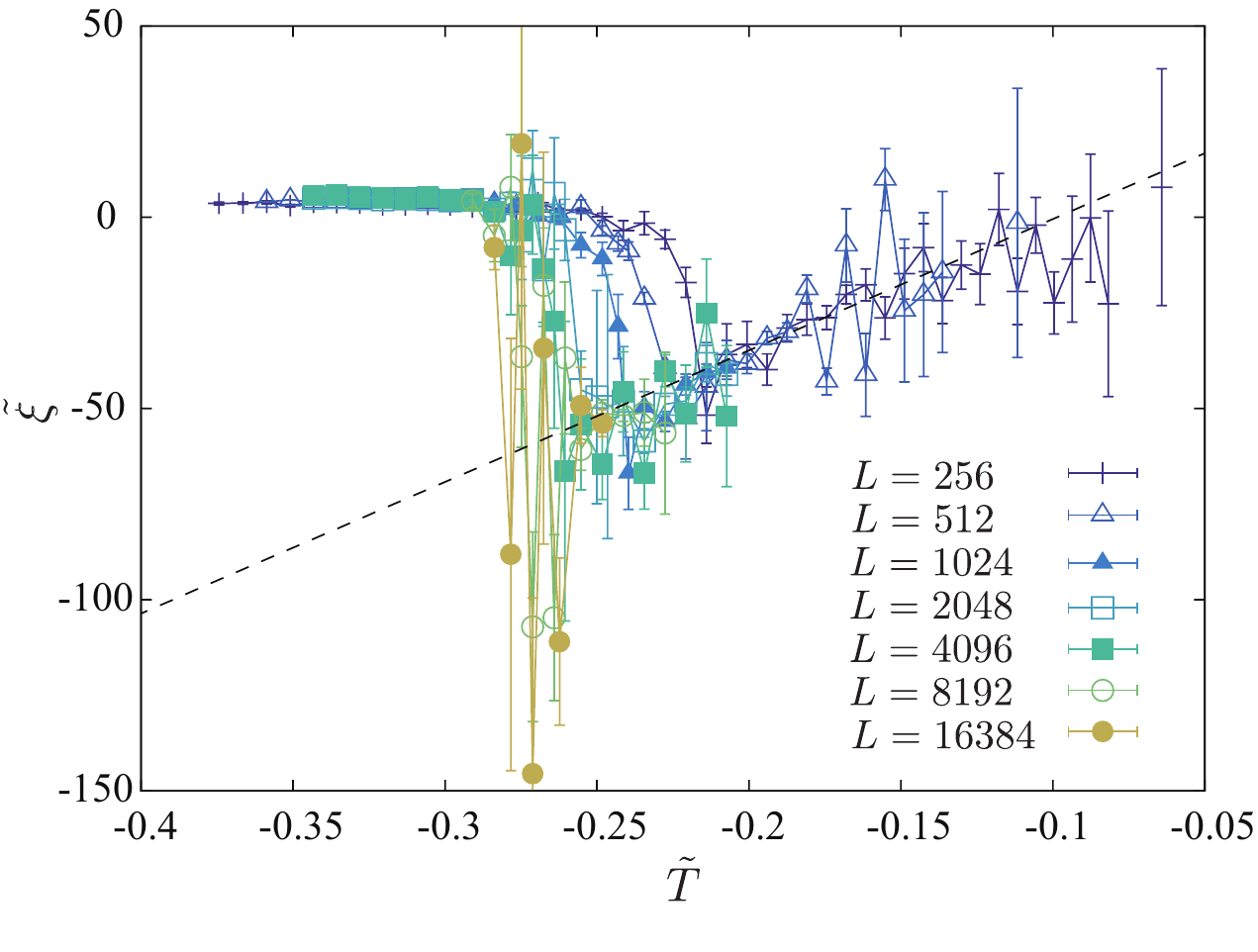}
 \end{center} 
 \caption{(Color online) Modified correlation length $\tilde{\xi}$ as a function of $\tilde{T}$ for various system
  sizes. The dashed line is for Eq.~\eqref{eq:linear_bc} with $\log b = 34$ and $c = 344$, which were obtained from the fitting to $L=256$ data at $-0.215 \le \tilde{T} \le -0.1$.}  \label{fig:corr_diff}
\end{figure}

To better understand the deviation from the temperature dependence in Eq.~\eqref{eq:no_Z2}, here we plot the correlation length in another way. Suppose the temperature dependence of the correlation length is described as
\begin{equation}
\xi = b T^c \exp\left(a/T\right).
\label{eq:correlation_abc}
\end{equation}
Then, from the two different temperatures $T_1$ and $T_2$, we define the modified correlation length and temperatures as 
\begin{align}
\tilde{\xi}(T_1, T_2) \equiv \frac{T_1 \log \xi(T_1) - T_2 \log \xi(T_2)}{T_1 - T_2} \notag \\
\tilde{T}(T_1, T_2) \equiv \frac{T_1 \log T_1 - T_2 \log T_2}{T_1 - T_2}.
\end{align}
By substituting Eq.~\eqref{eq:correlation_abc} to $\tilde{\xi}$ and $\tilde{T}$, we obtain the simple relation
\begin{equation}
\tilde{\xi} = \log{b} + c \tilde{T}.
\label{eq:linear_bc}
\end{equation}
In Fig.~\ref{fig:corr_diff}, we show $\tilde{\xi}$ as a function of $\tilde{T}$ by setting $T_1$ and $T_2$ to adjacent temperatures. As we decrease $\tilde{T}$, we see the clear system-size-dependent onset of the deviation from the linear behavior. Apart from this, as we argue below, we anticipate a system-size-independent crossover from the high-temperature linear behavior to the low-temperature behavior, which we may or may not be already seeing at around $\tilde{T} \simeq -0.25$ or $T/\tilde{J}\simeq 0.285$. Although the high-temperature linear behavior is consistent with Eq.~\eqref{eq:linear_bc}, the estimated exponent $c$ obtained from the fitting of $L=256$ data is unphysically large ($c = 344(51)$) compared with that estimated from the non-linear sigma model, Eq.~\eqref{eq:no_Z2}. This observation indicates a crossover from the observed high-temperature behavior with a large slope to the low-temperature behavior with a small slope consistent with Eq.~\eqref{eq:no_Z2}. Indeed, in Ref.~\citen{KawamuraYO2010}, a crossover of the correlation length between Eqs.~\eqref{eq:correlation} and \eqref{eq:no_Z2} was discussed by assuming a finite-temperature $Z_2$-vortex phase transition. In the scenario, the crossover temperature of the correlation length is identical to the phase transition temperature $T_v$. Although the observed behavior is consistent with this scenario, note that Fig.~\ref{fig:corr_diff} indicates a crossover even when we take the scenario of $T_v=0$,  because it is hardly imaginable that the extremely large effective value of the exponent $c$ survives in the low-temperature limit. Thus, our analysis supports a possibility of nontrivial correlation-length crossover in two-dimensional frustrated magnets, even if there is no true $Z_2$-vortex transition in the thermodynamic limit.
 
\begin{figure}[th] 
 \begin{center} 
  \includegraphics[width=0.9\linewidth]{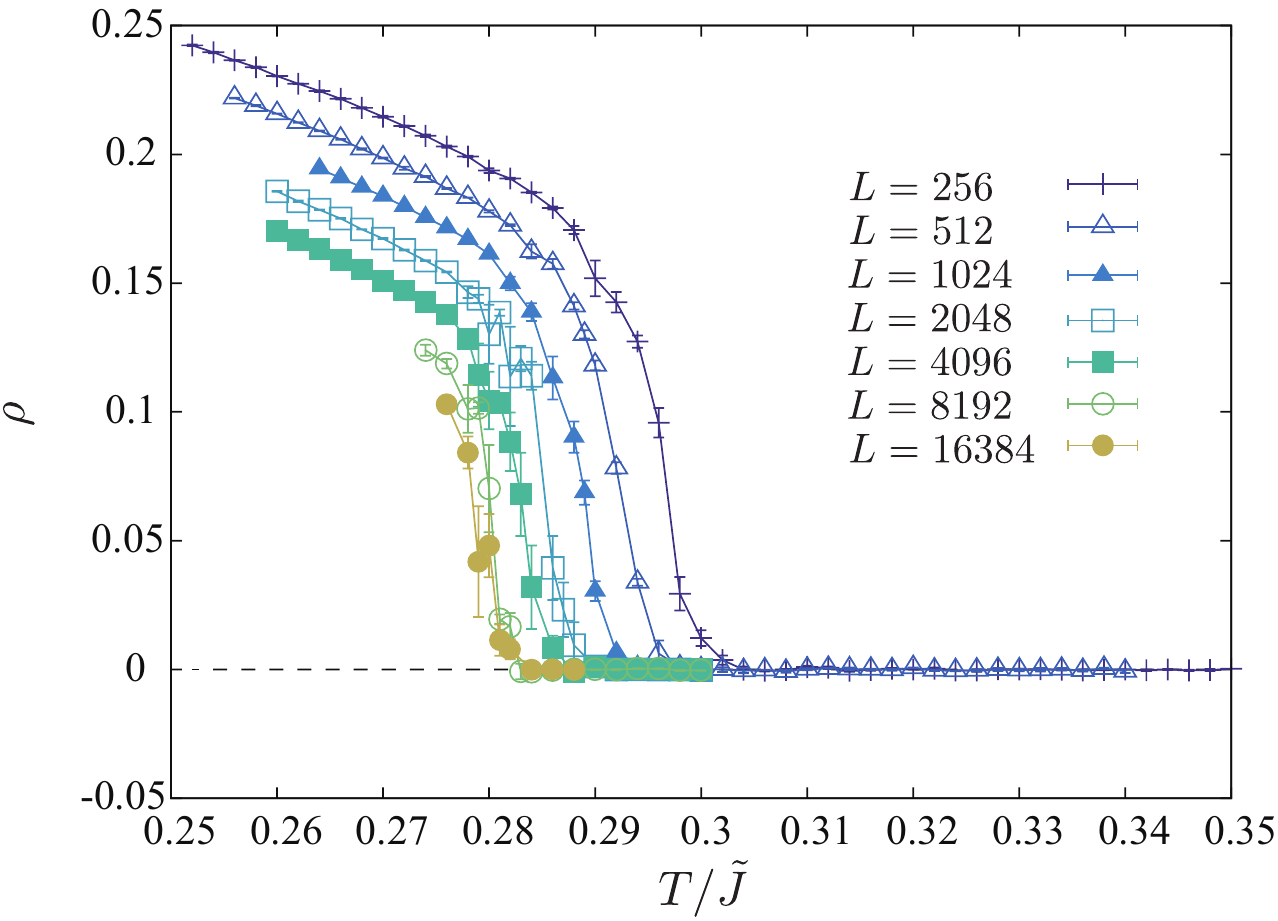}
 \end{center} 
 \caption{(Color online) Helicity modulus of $RP^3$ model for various system sizes.  The horizontal dashed line indicates $\rho = 0$.}  \label{fig:helicity}
\end{figure}

Next, to examin the separation of two length scales from another aspect, we also plot the model's helicity modulus $\rho$ for various system sizes in Fig.~\ref{fig:helicity}. Here, we define the helicity modulus in the space of the $\SO(3) \times \SO(3)$ model; we consider, \textit{e.g.}, the twist of $a$ and $b$ components around the $c$-axis. Because we expect a finite spin correlation length for $T < T_v$, the helicity modulus must be zero in sufficiently large systems. On the other hand, the vorticity modulus must be finite for $T < T_v$. Thus, when we observe a situation where the helicity modulus is zero and the vorticity modulus is finite in a certain temperature range, it can be another strong evidence of the $Z_2$-vortex transition. In the present MC simulations, however, we see that both the helicity modulus and the vorticity modulus are finite at low temperatures. This finding is consistent with the observation that the correlation length at $T_v$ is larger than $L=16384$.

\section{Conclusions}
\label{sec:conclusion} 
In this study, we investigated the nature of the possible topological phase transition, the $Z_2$-vortex transition, in a two-dimensional RP3 model. The model can be derived from the Heisenberg model by removing the intratriangle spin fluctuation and keeping only the degrees of freedom concerning the direction of the faces spanned by spins that locally have a planar order. Therefore, we believe that the frustrated Heisenberg model follows essentially the same scenario as the present model.  The most fundamental aspect of the $Z_2$-vortex transition, if it exists, is that the spin correlation length at the phase transition remains finite, and only the topological property, that is the existence of free vortices, changes through the phase transition. To clarify the existence of the $Z_2$-vortex transition, we considered a low-temperature effective model of the triangular lattice antiferromagnetic Heisenberg model. By keeping the topological property of the original model, we treated the $\SO(3) \times  \SO(3)$ model, which is identical to the $RP^3$ model, on the square lattice as the effective model for the $Z_2$-vortex transition.

By the parallel cluster MC simulation, we obtained equilibrium properties of the model up to $L=16384$, which is about ten times larger than the system sizes investigated in previous simulations of the triangular lattice Heisenberg model\cite{KawamuraM1984}. From the order parameter of the $Z_2$-vortex transition, we extracted a characteristic temperature that is expected to converge the transition temperature in the thermodynamic limit. We extrapolated the characteristic temperature assuming the mean-field exponent of the $Z_2$-vortex transition and estimated the transition temperature as $T_v/\tilde{J} \simeq 0.250$. We also confirmed that the fitting curve in this extrapolation is better than that assuming no $Z_2$-vortex transition.

We also calculated the correlation length at around $T_v$. One of the fundamental characteristics of the $Z_2$-vortex transition is the finite spin correlation length for $T \le T_v$; thus, it is important to confirm that the correlation length around the estimated $T_v$ is smaller than the system size treated in the present MC simulation. From our large-scale simulations, we found that even at the higher estimation of the transition temperature, $T_v/\tilde{J}=0.269$, the correlation length is larger than the maximum system size, $L=16384$, and much larger than that previously estimated in the triangular lattice Heisenberg model\cite{KawamuraYO2010}. Because the correlation length increases exponentially toward our best estimate $T_v/\tilde{J} \simeq 0.250$, the correlation length at the transition temperature is expected to be larger than $10^5$. Thus, we conclude that to observe the decoupling of the spin correlation length and the vortex correlation length, we should investigate systems with $ L > 10^5 $. Although it is too large to deal with in the near future, we note that the correlation length at the transition temperature is not a universal property. When we consider a different model, the correlation length at the $Z_2$-vortex transition can be much smaller, and we might investigate a clear topological phase transition for smaller system sizes. Searching for such systems could be a possible direction to investigate the nature of the $Z_2$-vortex transition.  

In addition, we showed the possibility of nontrivial crossover in the temperature dependence of correlation length. Although it can be understood from the finite-temperature $Z_2$-vortex transition\cite{KawamuraYO2010}, results of our analysis indicated that even if there is no true phase transition, the correlation length should show a crossover toward the low-temperature behavior described by the nonlinear sigma model. The existence of such a crossover means that $Z_2$ vortices play an important role in two-dimensional frustrated Heisenberg models even if they do not induce a true phase transition in the thermodynamics limit.

\section*{Acknowledgment}
We thank Hirotaka Kaneko for sharing with us the results of his preliminary calculation. The numerical calculations were performed on the K computer at RIKEN and supercomputers at ISSP, The University of Tokyo, through the support of the HPCI System Research Project (hp150116) and the Computational Materials Science Initiative (CMSI) supported by the Ministry of Education, Culture, Sports, Science, and Technology, Japan. The present work is financially supported by JSPS KAKENHI Nos.~15K17701, 19K03740, 22H01179, and 23H01092. T.O. acknowledges the support from the Endowed Project for Quantum Software Research and Education, The University of Tokyo (https://qsw. phys.s.u-tokyo.ac.jp/). 

\bibliographystyle{jpsj_mod}
\bibliography{RP3}

\begin{thebibliography}{10}

\bibitem{Diep2004}
{\em Frustrated Spin systems}, ed. H.~T. Diep (World Scientific Publishing,
  Singapore, 2004).

\bibitem{LacroixMM2011}
{\em Introduction to Frustrated Magnetism}, ed. C.~Lacroix, P.~Mendels, and
  F.~Mila (Springer, Berlin, Heidelberg, 2011).

\bibitem{Balents2010}
L.~Balents, Nature {\bfseries 464},  199 (2010).

\bibitem{LiaoXCLXHNX2017}
H.~J. Liao, Z.~Y. Xie, J.~Chen, Z.~Y. Liu, H.~D. Xie, R.~Z. Huang, B.~Normand,
  and T.~Xiang, Phys. Rev. Lett. {\bfseries 118},  137202 (2017).

\bibitem{GingrasMC2014}
M.~J.~P. Gingras and P.~A. McClarty, Rep. Prog. Phys. {\bfseries 77},  056501
  (2014).

\bibitem{Berezinsky1970}
V.~L. Berezinsky, Sov. Phys. JETP {\bfseries 32},  493 (1971).

\bibitem{KosterlitzT1973}
J.~M. Kosterlitz and D.~J. Thouless, J. Phys. C: Solid State Phys. {\bfseries
  6},  1181 (1973).

\bibitem{KawamuraM1984}
H.~Kawamura and S.~Miyashita, J. Phys. Soc. Jpn. {\bfseries 53},  4138 (1984).

\bibitem{KawamuraYO2010}
H.~Kawamura, A.~Yamamoto, and T.~Okubo, J. Phys. Soc. Jpn. {\bfseries 79},
  023701 (2010).

\bibitem{OkuboK2010}
T.~Okubo and H.~Kawamura, J. Phys. Soc. Jpn. {\bfseries 79},  084706 (2010).

\bibitem{KunzZ1989}
H.~Kunz and G.~Zumbach, J. Phys. A: Math. Gen. {\bfseries 22},  L1043 (1989).

\bibitem{KunzZ1992}
H.~Kunz and G.~Zumbach, Phys. Rev. B {\bfseries 46},  662 (1992).

\bibitem{LebwohlL1972}
P.~A. Lebwohl and G.~Lasher, Phys. Rev. A {\bfseries 6},  426 (1972).

\bibitem{Tomita2014}
Y.~Tomita, Phys. Rev. E {\bfseries 90},  032109 (2014).

\bibitem{LathaS2018}
B.~K. Latha and V.~S.~S. Sastry, Phys. Rev. Lett. {\bfseries 121},  217801
  (2018).

\bibitem{SouthernX1995}
B.~Southern and H.-J. Xu, Phys. Rev. B {\bfseries 52},  R3836 (1995).

\bibitem{WintelEA1995}
M.~Wintel, H.~U. Everts, and W.~Apel, Phys. Rev. B {\bfseries 52},  13480
  (1995).

\bibitem{CaffarelADM2001}
M.~Caffarel, P.~Azaria, B.~Delamotte, and D.~Mouhanna, Phys. Rev. B {\bfseries
  64},  014412 (2001).

\bibitem{KawamuraK1993}
H.~Kawamura and M.~Kikuchi, Phys. Rev. B {\bfseries 47},  1134 (1993).

\bibitem{SwendsenW1987}
R.~H. Swendsen and J.-S. Wang, Phys. Rev. Lett. {\bfseries 58},  86 (1987).

\bibitem{KawamuraY2007}
H.~Kawamura and A.~Yamamoto, J. Phys. Soc. Jpn. {\bfseries 76},  073704 (2007).

\bibitem{SouthernY1993}
B.~W. Southern and A.~P. Young, Phys. Rev. B {\bfseries 48},  13170 (1993).

\bibitem{AzariaDM1992}
P.~Azaria, B.~Delamotte, and D.~Mouhanna, Phys. Rev. Lett. {\bfseries 68},
  1762 (1992).

\end{thebibliography}
\end{document}